\documentclass{article}
\usepackage[utf8]{inputenc}
\usepackage{authblk}
\providecommand{\keywords}[1]{\textbf{\textit{Keywords:}} #1}
\providecommand{\PACS}[1]{\textbf{PACS} #1}

\usepackage{amsmath}
\usepackage{graphicx}
\graphicspath{ {./images/} }
\usepackage{hyphenat}
\usepackage{latexsym}
\usepackage{stackrel,amssymb}
\usepackage{esint}

\title{Search for gravitational-neutrino correlations on ground-based detectors}
\author[1]{Svetlana Andrusenko}
\author[2,3]{Yurii  Gavriluk}
\author[2]{Andrei Gusev}
\author[2,4]{ Daniil Krichevskiy}
\author[2]{Sergei  Oreshkin}
\author[2]{Sergei  Popov}
\author[2,3]{Valentin Rudenko \thanks{Corresponding Author:  valentin.rudenko@gmail.com}}

\affil[1]{Physics Department, Bauman Moscow State Technical University}
\affil[2]{Sternberg Astronomical Institute, Lomonosov Moscow State University}
\affil[3]{Institute for Nuclear Research, Russian Academy of Sciences}
\affil[4]{Faculty of Science, University of Bern}
\date{2022}

\begin{document}

\maketitle

\begin{abstract}
    The problem of joint data processing from ground-based gravitational and neutrino detectors is considered in order to increase the detection efficiency of collapsing objects in the Galaxy.  The development of the "neutrino - gravitational correlation" algorithm is carried out within the framework of the theory of optimal filtration as applied to the well-known OGRAN and BUST facilities located at the BNO INR RAS.  The experience of analyzing neutrino and gravitational data obtained during the outburst of supernova SN1987A is used.  Sequential steps of the algorithm are presented, formulas for estimating the statistical efficiency of a two-channel recorder are obtained.
\end{abstract}

\keywords{gravitational waves; neutrino telescope; multi-messenger astronomy}

\PACS{04.30.w; 04.80.Nn; 97.60.Bw}

\section{Introduction}

After the registration by ground-based gravitational-wave (GW) interferometers LIGO and Virgo of bursts of gravitational radiation from the merger of relativistic binaries: - black holes  \cite{PhysRevLett.116.061102,PhysRevLett.118.221101} and neutron stars \cite{PhysRevLett.119.161101}, as well as after the discovery of a relativistic object simultaneously generating gravitational and electromagnetic (gamma) signals, with  subsequent optical afterglow \cite{PhysRevLett.119.161101}, one can confidently speak about the emergence of a new GW channel of astrophysical information.  This new channel, effectively supplementing the previously developed methods for registering cosmic rays and particle fluxes, contributes to the development of a new heuristic approach to astronomical observations, called "multi-messenger astronomy" or in Russian transcription - "multi-channel astronomy".

 To date, there are no operating large GW interferometers in Russia, although a project to create such an installation in Siberia is known \cite{universe6090140,Rudenko2021}.  The only gravitational detector in the kilohertz frequency range is the combined optoacoustic antenna OGRAN, developed jointly by the Russian Academy of Sciences and Moscow State University.  Details of the structure, principle of operation, and technical parameters of this setup can be found in \cite{doi:10.1063/1.4883901,Rudenko2017}.  The latest upgraded version, along with the current experimental characteristics, is presented in \cite{Rudenko2020}.  The proposed article is devoted to the analysis of the data processing technique of this antenna, for which it is first useful to briefly recall its principles and design scheme.
 
 OGRAN is an installation located in the underground premises of the Baksan Neutrino Observatory of the Institute of Nuclear Research, Russian Academy of Sciences, designed to search for collapsing stars in the Galaxy together with the BUST neutrino telescope.  Both instruments are sensitive enough to register collapses in our Galaxy as rare events with an average rate of $0.03$ events per year.
 
\begin{figure}[h]
\includegraphics[width=12 cm]{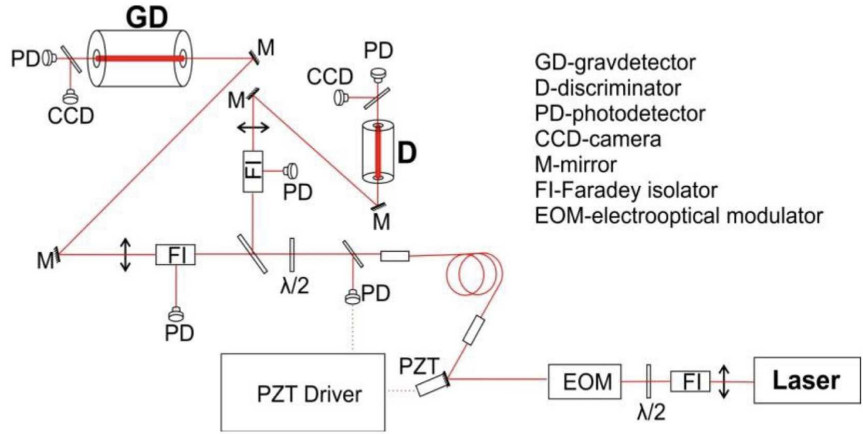}
\caption{ Principle scheme of the OGRAN antenna \label{fig1}}
\end{figure}

 Structurally, OGRAN (Figure~\ref{fig1}) consists of two arms (or "feedback loops").  The first one, the “detector arm”, contains a large aluminum bar - an acoustic resonator of longitudinal vibrations with a mass $M\approx 2\ t$ and a length  $L\approx 2\ m$.  The second, “discriminator arm”, is equipped with a small glass-ceramic bar ($M\approx 30\ kg$, $L\approx 40\ cm$).  Both bars have a cylindrical shape with drilled central axial channels for a light guide.  Pairs of mirrors are mechanically attached to the ends of the bars, forming Fabry-Perot (FP) standards.  When illuminated from a common (tunable) laser with a wavelength of $\lambda \approx  1\ \mu m$, these FP standards form the optical degree of freedom of the OGRAN antenna, with which GW can interact directly, i.e. acting on the light field in the FP.  Antenna tuning in the operating mode binds the laser frequency to the selected optical FP resonance in the detector arm.  In this case, information about the acoustic oscillations of the detector (large bar) is encoded in the frequency of the optical pump.  In the arm of the discriminator (small bar), tuning to optical resonance is carried out by moving one of the FP mirrors due to feedback circuits at low (quasi-static) frequencies.  As a result, frequency demodulation is performed and the output signal reproduces the acoustic vibrations of the gravitational detector.  In general, the design of the OGRAN antenna is a differential circuit known as the “comparator of optical standards” \cite{Rosa_2002,Bezrukov2010}.
 
\begin{figure}[h]
\includegraphics[width=12 cm]{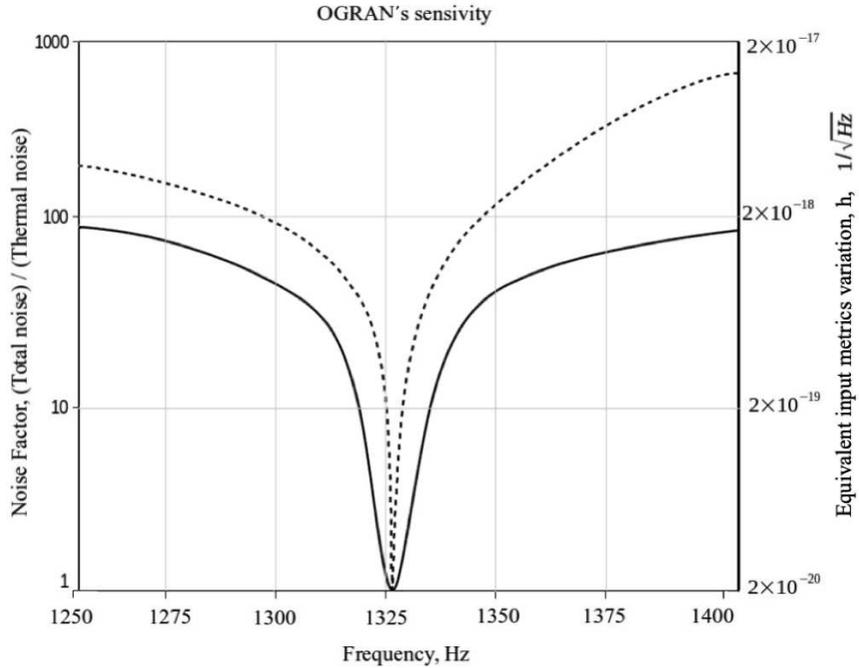}
\caption{ Noise spectral density of the OGRAN antenna (from \cite{doi:10.1142/S0217751X20400072}). Doted line - the experimental curve of old
version of the setup $F \sim 2000$; normal line - the curve of modernized setup $F \sim 30000$. 
\label{fig2}}
\end{figure}

 The upgraded OGRAN version has the following optical parameters.  Optical pump power in each of the arms is $\sim 200\ mW$, FP sharpness (finesse)in the detector arm is 30,000 and 70,000 in the discriminator arm.  Interference contrast $0.3\div 0.7$.
 The sensitivity curve of the antenna is shown in Figure~\ref{fig2}. It can be seen that for the current OGRAN version those metric variations $h$ are registered which cause deformation perturbations of the length of the detector (large bar)
 $h = (\bigtriangleup L/L)_{f\  }\sim 1\cdot 10^{-18} \ Hz^{-1/2}$ in  $\bigtriangleup f\sim 30 \ Hz$ bandwidth or $(\bigtriangleup L/L)_{f\  }\sim 1\cdot 10^{-19} \ Hz^{-1/2}$ in $\bigtriangleup f\sim 4 \ Hz$ bandwidth around $1.3 \ kHz$ acoustic resonance center frequency. 
 
 It should be noted that this sensitivity is not enough to detect signals from collapsing stars at a distance of about $8 \ kpc$ (the center of the Galaxy). An increase in sensitivity by two orders of magnitude without changing the design of the antenna is, in principle, possible by forcing its parameters: sharpness (finesse) of FP resonators in the arms up to $300000$ and optical pump power up to $20 \ W$. In this case, the sensitivity will reach the level of $10^{-22} \ Hz^{-1/2}$ in a band of about $100 \ Hz$ \cite{doi:10.1063/1.4883901,Rudenko2017,Rudenko2020}.
 
  We also note that OGRAN is located in the underground premises of the BNO, not far from the BUST neutrino detector (BUST) which plays an important role in the search for and detection of collapsing objects within the Galaxy.  The joint coordinated use of these tools increases the likelihood of success in this task.
 The goal of this study is to develop an efficient algorithm for joint analysis (processing) of data from both detectors - neutrino and gravitational.

\section{ Mathematical model of the gravitational detector and GW signal}

The acoustic part of the OGRAN detector is the same as that of resonant solid-state detectors \cite{Astone2007,Astone_2002}, i.e. appears to be the main antisymmetric mode of longitudinal oscillations of a cylindrical
bar (with length $L$ much greater than its radius $r$), which makes it possible to model the detector as a one-dimensional oscillator with two parameters: normal mode period $\tau_{0}$  and relaxation time $\tau_{r}$ .

As a generalized model of the GW signal (external force impact on the detector), a short  wave train (or radio pulse) with amplitude $F_{s}$ and duration $\tau_{s} \ll  \tau_{r}$ filled with oscillations close to its resonant frequency $\omega_{0}$, is considered. In accordance with the GR formulas, in the weak field approximation, the action of a monochromatic gravitational wave (GW) with frequency $\omega$ on a one-dimensional detector is equivalent to a quasi-resonant swinging force with an amplitude $F_{s}=0.5m\omega ^{2}Lh_{0}$, where $m$ is the equivalent mass of the detector, $h_{0}$ is the dimensionless amplitude of the metric variations of the transferred GW associated with the gravitational energy flux density in the wave \cite{BicakRudenko,Pizella2016}. Since due to the equivalence principle, the action of the GW on the detector is determined by the transferred acceleration, regardless of its mass, it is natural to introduce the notation $f_{s}=F_{s}/m $.
In the mathematical formulation of the problem of calculating the effect of a GW train on a gravitational detector, we use the apparatus of complex envelopes \cite{Gonorovsky}. Thus we represent $f_{s}(t)=Re[\tilde{f}_{s} (t)e^{j\omega_{0} t}]$. For a  wave train, the form of the complex envelope depending on time is

\begin{equation}\label{eq1}
\tilde{f}_{s} (t)=\begin{cases}A,&0\leqslant t\leqslant \hat{\tau } \\ 0,& t>\hat{\tau }   \end{cases},
\end{equation}
where the A is a  wave train amplitude and  $ \hat{\tau} $  -  wave train duration. 

To describe the thermal (Brownian) oscillations of the detector, we introduce $f_{T} (t)$ - the random Nyquist force, i.e. white noise with a spectral intensity $N_{0}=4k_{B}\Theta  m\delta {}$, which depends on the absolute temperature $\Theta $, and the value of the losses of the detector given by the attenuation factor $\delta =\tau_{r}^{-1}$, associated with the quality factor of the detector $Q=\omega _{0}/(2\delta ) \gg 1$ (the so-called "Langevin approach" to description of Brownian motion \cite{Radiophysics} ).
As a result, the response spectrum of the acoustic detector (the resonant unit of the OGRAN setup) is obtained by passing the GW train $f_{s}(t)$ through the frequency transfer function of the oscillator
\begin{equation}\label{eq2}
g(j\omega )=\frac{1}{m\left( \omega ^{2}_{0}-\omega ^{2}+2j\delta \omega \right)  },
\end{equation}
The time form of the response $s(t)$ is determined by the impulse response $g(t)$ through the convolution transformation
\begin{equation}\label{eq3}
s(t)=f_{s}(t)\ast g(t)=\int f_{s}(\tau )g\left( t-\tau \right)  d\tau \leftrightarrow S\left( j\omega \right)  =f_{s}(j\omega )g(j\omega ),
\end{equation}
where the sign  $\leftrightarrow$ means the transition from the temporal form of the response to the spectral form, which reduces to the product of the spectra of the considered functions. We immediately present the form of the spectrum $f_{s}(j\omega ) $ of the complex envelope $\tilde{f}_{s} (t)$ of the signal  wave train, which is used in further analysis

\begin{equation}\label{eq4}
{f}_{s} (j\omega )=\frac{A}{j\omega } (1-e^{-j\omega \hat{\tau } }).
\end{equation}

In addition to the signal $f_{s}$ and thermal $f_{T}$ effects on the detector, one should take into account the “measuring noise” (or sensor noise), which for OGRAN is associated with the optoelectronic conversion of the acoustic oscillations of the bar into a measured electrical signal. Usually fluctuations of measuring instruments are represented as an additive interference $\xi (t)$ of the white noise type in a limited frequency band at a spectral intensity $N_{\xi }$, the estimate of which is taken from the experiment, $<\xi (t)\xi (t+\tau )>=N_{\xi }\delta (\tau )$. Thus, roughly (without fine details) the output of an opto-acoustic gravitational detector can be written as

\begin{equation}\label{eq5}
y(t) = s(t) + n(t) .
\end{equation}

As indicated above, the signal part is represented by the convolution $s(t)=f_{s}(t)\ast g(t)$, the fluctuation part is formed as the sum $n(t)=\eta (t)+\xi (t)$, including thermal noise $\eta (t)=f_{T}(t)\ast g(t)$ with spectral density $N_{T}(\omega )=N_{0}\mid g\left(j\omega \right) \mid^{2}$ and measuring noise $\xi (t)$ . Considering that noises $\eta (t)$ and $\xi (t)$  are independent, the spectral intensity of the output of the detector decomposes into the sum of the spectral densities of thermal and additive noise $N(\omega )=N_{T}(\omega )+N_{\xi }$.

\section{ Optimal filtering algorithm}

The procedure for optimal filtering of the detector output of an optoacoustic detector $y(t) = s(t) + n(t)$ lies in passing it through a linear optimal filter with a transfer function

\begin{equation}\label{eq6}
K_{opt}(j\omega )=const\frac{S^{\ast }(j\omega )}{N(\omega )} e^{-j\omega t_{0}},
\end{equation}
where $\ast $ means complex conjugation, $t_{0}$ is the time of maximum response, $const$ means some arbitrary  constant. At the filter output, we get a narrow-band process $z(t)=Re[\tilde{z} (t)e^{j\omega_{0} t}]$  with a complex envelope $\tilde{z} (t)$.

To explain the physical meaning of the filtering $K_{opt}(j\omega )$, it is convenient to split it into two successive transformations $K_{opt}(j\omega )=K_{0}(j\omega )K_{1}(j\omega )$, thus separating the stage of the so-called “Wiener filtering” $K_{0}(j\omega )$ \cite{Pizella2016,Gonorovsky} from the subsequent matched filter $K_{1}(j\omega )$ \cite{Gonorovsky}. This is a consequence of the fact that in the problem under consideration there are two types of noise: white additive measurement noise and thermal noise, which has a finite spectrum in a given limited frequency band.
As a result, we arrive at a representation in which the transfer function of the Wiener filter for the total noise with an effective bandwidth $\Omega \sim \omega_{0} \left( N_{0}/4m^{2}N_{\xi }\right)^{1/2}   $ reads as
\begin{equation}\label{eq7}
K_{0}(j\omega )=\frac{N_{T}(\omega )}{N(\omega )} =\frac{N_{0}\mid g(\omega )\mid^{2} }{N_{0}\mid g(\omega )\mid^{2} +N_{\xi }}
\end{equation}
and the transfer function of the matched filter against the background of thermal noise reads as
\begin{align}\label{eq8}
\begin{gathered}
K_{1}(j\omega )=\frac{K_{opt}(j\omega )}{K_{0}(j\omega )} =const\frac{S^{\ast }(j\omega )}{N_{0}\left| g\left( j\omega \right)  \right|^{2}  } e^{-j\omega t_{0}}= \\ const\frac{f^{\ast }_{s}(j\omega )g^{\ast }\left( j\omega \right)  }{N_{0}\mid g(j\omega )\mid^{2} } e^{-j\omega t_{0}}=const\frac{f^{\ast }_{s}(j\omega )}{N_{0}g(j\omega )} e^{-j\omega t_{0}}. 
\end{gathered}
\end{align}
With small losses in the vicinity of the resonant frequency $\omega _{0}\pm \omega $, $ \omega  \ll \omega _{0}$, the formula for $ K_{1}$ can be simplified in accordance with the approximation

\begin{equation}\label{eq9}
g(\omega_{0} +\omega )\approx \left[ m\left( \delta +j\omega \right)  \right]^{-1}   
\end{equation}

Then, collecting formulas (\ref{eq4}, \ref{eq8}, \ref{eq9}), we get a clearer form of the transfer function of the matched filter $K_{1}(\omega)$;
Considering that $\tilde{f}_{s} (j\omega)\approx (1/2)f_{s}(j\left( \omega_{0}+\omega\right)  )$ we come to the form
\begin{equation}\label{eq10}
K_{1}(j\left( \omega_{0}+\omega\right)  ) \approx const\left( 1-e^{j\omega\hat{\tau } }\right), \ \text{for} \ \delta \rightarrow 0
\end{equation}
which in time domain corresponds to the operation of the difference link
for the variable $x\left( t\right)  \longleftrightarrow x\left( j\omega\right)  $ at the output of the Wiener filter. As a result, the procedure for optimal processing of the OGRAN detector output $y(t)$ is the chain of transformations

\begin{equation}\label{eq11}
y(t)\leftrightarrow y(j\omega)\rightarrow K_{0}(j\omega)\rightarrow x(j\omega)\rightarrow K_{1}(j\omega)\rightarrow z(j\omega)\leftrightarrow z(t)
\end{equation}

The ratio between the spectra $x\left( j\omega\right)$ and $z\left( j\omega\right)$ near the resonant frequency $\omega_{0}$ is determined by the condition $ 2x\left( \omega+\omega_{0}\right)  K_{1}\left( \omega+\omega_{0}\right)  =z\left( \omega+\omega_{0}\right), \mid \omega\mid \ll \omega_{0}$. The role of the Wiener filter lies in the frequency cutoff of the background of the measuring white noise $\xi (t)$. The subsequent matched filtering $K_{1}(j\omega)$ is determined by the spectrum of thermal fluctuations $N_{0}\left| g\left( j\omega\right)  \right|^{2}  $ and, under the condition of short GW signals $\widehat{\tau } \ll \tau_{r} $ , is reduced to the final procedure of the difference link $\widetilde{x} (t)-\widetilde{x} (t-\widehat{t} )=\widetilde{z} (t)$  for the complex envelope. Note that in practice such an envelope can be constructed in terms of the quadrature components of the output $z(t)$.

Summarizing the procedure for optimal processing of the OGRAN output, we write out in explicit form the expression for the transfer function of the optimal filter $K_{opt}(j\omega)$ in the vicinity of the resonant frequency $\omega_{0}$ and the formula for calculating the signal-to-noise ratio (SNR) at its output, namely
\begin{align}\label{eq12}
K_{opt} (j\left( \omega_{0}+\omega\right)) \approx const\left( 1-e^{j\omega\hat{\tau } }\right) \frac{\Omega^{2}}{\Omega^{2} +\omega^{2}_{0}}   
\end{align}

\begin{equation}\label{eq13}
\begin{gathered}
SNR=q=\left( 1/\pi \right)  \int\limits^{\infty }_{0} \frac{\left| S\left( j\omega\right)  \right|^{2}  }{\left( N_{0}\left| g\left( j\omega\right)  \right|^{2}  +N_{\xi }\right)  } d\omega = \\ \left( 1/4\pi \right)  \int\limits^{\infty }_{-\infty } \frac{\left| f_{s}\left( j\omega\right)  \right|^{2}  }{\left( N_{0}\left| g\left( j\omega\right)  \right|^{2}  +N_{\xi }\left| g\left( j\omega\right)  \right|^{-2}  \right)  } d\omega = \\ \left( \frac{A^{2}\widehat{\tau } }{2N_{0}} \right)  \left\{ 1-\left( \frac{1}{\Omega \widehat{\tau } } \right)  \left( 1-e^{-j\Omega \widehat{\tau } }\right)  \right\} 
\end{gathered}
\end{equation}

here $\Omega$ is the previously introduced denotation of the Wiener filter effective bandwidth. 

It is interesting to compare $SNR=q$ for two signal models: the  wave train model adopted by us, for which the GW burst spectrum width is less than the Wiener filter bandwidth , i.e. $\Omega \widehat{\tau } >1$, and a very short burst like $\delta \left( \tau \right)  $ – an impulse for which $\Omega \widehat{\tau } \ll 1$ (the model used for Italian cryogenic antennas \cite{Astone2007,Astone_2002,Pizella2016}). In the first case, from (\ref{eq13}) we obtain $q_{1}\approx \left( A^{2}\widehat{\tau } /2N_{0}\right)  $, while in the second case we have a much smaller value $q_{2}=q_{1}\Omega \widehat{\tau } /2$.

\section{Envelope outliers detections}

The optimal filtering procedure of the OGRAN output signal described above, in fact, leads the operator to the need to observe bursts or outliers at the output of the difference link
$\triangle \widetilde{x} (t,\tau )=\widetilde{z} (t)$ constructed taking into account the phase of the quasi-harmonic variable $z(t)$. In the absence of a priori information about the moment of arrival of the GW signal and any information from parallel recording channels (multi-messenger astronomy), the only method for detecting external influences is the so-called Neyman-Pearson strategy (NPS) \cite{1975MIzSR....T....L}. In this methodology, an indicator of the presence of an impact is the excess of the number of emissions above the average for the selected threshold level $C_{\alpha }$, which is determined by the statistics of the observed variable.

For sufficiently high threshold levels $\left( C/\sigma \right)  >1$, where $\sigma$ is  the standard deviation, random outliers can be considered independent and subject to the Poisson distribution, representing the probability $P\left\{ k\right\}  $ of the number of outliers $k$ that occurred during the observation time $T$. In the simplest case, the average number of outliers $\lambda $ is assumed to be constant thus a flow of uniform intensity is considered. Then the “Poisson law” is written as $P\left\{ k\right\}  =\lambda^{k} (1/k!)e^{-\lambda }$, where $\lambda =<n\left( C,T\right)  >$ ($< >$ sign of statistical averaging).
     
At the OGRAN output (after the optimal filter) there is a narrow-band Gaussian random process $z(t)=Re[\tilde{z} (t)e^{j\omega_{0} t}]$ with the correlation function $k(t)=<z(t)z(t+\tau )>=Re\left[ \tilde{k} (\tau )e^{j\omega_{0} \tau }\right]$ , where $\tilde{k} (\tau )=\sigma^{2} \rho (\tau )$) is a complex envelope with variance $\sigma^{2} =<z^{2}(t)>$ and correlation coefficient $\rho \left( \tau \right)$ . Accordingly, for the average number of envelope outliers $\left| \widetilde{z} \left( t\right)  \right|  $ above the threshold level $C$ during the time $(0,T)$ the following formula takes place \cite{Radiophysics,1975MIzSR....T....L,Outliers}
\begin{align}\label{eq14}
<n(c,T)>=T\left( -\frac{{\rho }^{\prime \prime }_{0}  }{2\pi } \right)^{1/2}  \frac{C}{\sigma } e^{-\frac{C^{2}}{\sigma^{2} } },
\end{align}
where ${\rho }^{\prime \prime }_{0}  =\frac{{}d^{2}\rho (\tau )}{d\tau^{2} } $ at  $\tau=0$.

The application of NPS to the “observed variable” (in this case, to outliers at the output of the difference link $\triangle \widetilde{x} (t,\tau )$ - at the first step, it consists in calculating the value of the threshold level $C_{\alpha }$ corresponding to the selected statistical error of the 1st kind $\alpha <1$, - the so-called "probability of false alarm" in the theory of detection, or "probability of chance" in nuclear physics (the mathematical term "significance level" is also used). For a Poisson distribution with an average value (\ref{eq14}) it is not difficult to find that the threshold
\begin{align}\label{eq15}
C^{2}_{\alpha }=2\sigma^{2} ln\left[ \left( -\frac{{\rho }^{\prime \prime }_{0}}{2\pi } \right)^{1/2}  \frac{1}{\alpha } \right]  
\end{align}
corresponds to the condition for the absence of outliers $P\left\{ k=0\right\}  =1-\alpha$. There are no outliers above such a threshold with an error $\alpha$.
To specify formulas (\ref{eq14}), (\ref{eq15}), it remains to give the results of calculating the parameters ${\rho }^{\prime \prime }_{0}$ and $\sigma^{2} $
\begin{align}\label{eq16}
\sigma^{2} =B\left( \frac{\widehat{\tau } }{4\Omega^{2} } \right)  \left[ 1-\left( \frac{1}{\Omega \widehat{\tau } } \right)  \left( 1-e^{-\Omega \hat{\tau } }\right)  \right]  , 
\end{align}
\begin{align}\label{eq16.1}
-{\rho }^{\prime \prime }_{}  =\left( \frac{\Omega }{\widehat{\tau } } \right)  \left( 1-e^{-\Omega \widehat{\tau } }\right)  \left[ 1-\left( \frac{1}{\Omega \widehat{\tau } } \right)  \left( 1-e^{-\Omega \widehat{\tau } }\right)  \right]^{-1}  .
\end{align}

Within the framework of NPS, two approaches are possible to make a decision about an external influence (presence of a signal).

The first one (the standard approach) is tracking a random variable, the number of outliers during the observation time $T$, over a certain threshold level $\left( C/\sigma \right) $, the height of which is limited only by the independence condition of the outliers (the absence of their correlation). Then the excess of the observed number of releases over the average value, corresponding to high reliability of the event (standard - $0.95$) will serve as a statistical criterion for registering an external impact. This corresponds to the usual rule of accepting the "non-null hypothesis" i.e. hypotheses of "the presence of an external perturbation" that changes its own statistics of the observed variable.

The second approach ("absolute maximum criterion") is the observation at a high threshold level given by the condition (\ref{eq15}) of a single "excess event", i.e. the main (largest) maximum crosses the threshold level (\ref{eq15}) at least once.
Obviously, the first approach should in principle be more sensitive to weak influences, since it allows operation at relatively low levels. However, the degree of registration reliability (confidence limit) is expected to be higher in the second case. In general, the choice of approach depends on the physics of a particular problem.
To conclude this section, we present a formula for the "probability of correct detection" $D$ when using the "absolute maximum criterion" (second approach). It is not difficult to find that with a known ratio $SNR=q >>1$, the estimate of this indicator is
$D\approx \Phi \left( \sqrt{q} -\frac{C_{\alpha }}{\sigma } \right)  $, where $\Phi \left( x\right)  $ is the error function.

\section{ BUST and neutrino detection technique}

In the BNO, the registration of neutrino signals accompanying the appearance of collapsing stars in our Galaxy is included in the program of observations of the Baksan Underground Scintillation Telescope (BUST). This large-sized detector with dimensions $(17 \times  17 \times  11) \ m^{3}$ is located in an underground laboratory protected from cosmic particle flows by a rock corresponding to 850 meters of water equivalent \cite{Novoseltsev2017}. Structurally, the BUST consists of four horizontal and four vertical planes filled with cells (counters) of liquid scintillators. The total number of counters is 3184 with a total scintillator mass of 330 tons. Each counter is an aluminum container $(0.7 \times  0.7 \times  0.3) \ m^{3}$ filled with white spirit ($C_{n}H_{2n+2};n\approx 9$). However, in the "collapse program" only a part of the most protected (internal) counters is used, - the number of 1200 with a mass
$130 \ t$ targets. Thanks to improved protection, they have a relatively low background event count rate: $f=0.02 \ s^{-1}$
Most of the events that the BUST registers from the supernova (SN) explosion are inverse beta decay reactions: $\bar{\nu }_{e} +p\longrightarrow n+e^{+}$. (interaction of an electron antineutrino with a target proton generates a neutron and a positron). At a typical average antineutrino energy $E_{\nu e}=\left( 12-15\right) \ MeV$ \cite{1998PPN....29..254A}, the range of the produced positron $e^{+}$ will, as a rule, be contained in the volume of an individual counter. Therefore, the signal from SN will appear as a series of events, when only one counter is activated from the total number of cells on the installation (“single event”)
Thus, the strategy for searching for a neutrino burst at the BUST is to register a cluster (group) of single events during the time interval $\tau =20\  s$ (estimation of the duration of a neutrino burst from SN ).
For SN at a distance of $10 \ kpc$, the total energy emitted in a neutrino is $3\cdot 10^{53} \ erg$. With a target mass of 130 tons (three lower BUST planes), the estimate  of the expected number of events from a single collapse is $N\sim 35$ \cite{Novoseltsev2017,1998PPN....29..254A}. Of course, there is a random background of such events, created by the radioactivity of the environment, cosmic ray muons, false alarms of counters, etc. The background is such that it creates a cluster of $k = 8$ single events with a frequency of $0.138 \ year^{-1}$. At the same time, no more than 2 such clusters can be expected to appear in 10 years. The rate of formation of clusters from $k = 9$ background events is already much lower and amounts to $7\cdot 10^{-3} \ year^{-1}$. It follows that clusters with $k > 9$ cannot be created by the background and, therefore, are candidates for registering the SN event.

Identification of suspicious 20-second intervals in experimental data can be done in two ways. In the first one, a sliding 20-second time interval moves in discrete steps from one single event to the next so that there is always at least one event in the cluster (at the beginning of the interval). In this case, the Poisson distribution law of events may be violated. Another processing option is possible when the event clusters do not overlap. The beginning of each time interval coincides with the end of the previous one. The first interval is chosen arbitrarily. In this case, the distribution of clusters according to the number of events is strictly Poisson, but a cluster with a higher multiplicity can be lost due to some of the events falling into a neighboring cluster. The work \cite{Novoseltsev2017} refers to the use of both methods for mutual control. It also states that the "radius of sensitivity" of the BUST is approximately $20 \ kpc$. This region includes about 95 \% of the stars in our Galaxy. For more distant SNe, the number of single events in a cluster will be less than nine, in which case it is necessary to investigate correlations with other installations.
For the period from 1980 to 2021, the net observation time for collapses at the BUST was $\sim 36$ years which is the longest time of observation of the Galaxy on the same setup. During this time, not a single event of a candidate for the collapse of a stellar object (a cluster with $k\geq 9$) was registered. This leads to the value of the upper limit of the average frequency of gravitational collapses in the Galaxy $f_{col}<0.064\  year^{-1}$ at the 90 \% confidence level \cite{Novoseltsev2022}. Theoretical estimates of the frequency of occurrence of galactic SNe with core collapse give a value of approximately 2–5 events per century \cite{Adams_2013}.

\section{Algorithm for searching for neutrino-gravitational correlations}

Let us refine the form of the expected neutrino signal that accompanies the process of collapse of a relativistic star whose mass exceeds the critical one. According to \cite{1987ApJ...318..288M,Melson_2015}, the most active phase of the process develops in a short time of the order of $\sim 1\ s$. The neutrino luminosity curve usually has a large initial peak followed by weaker peaks reflecting radiation during bounces in the process of monotonic compression \cite{1987ApJ...318..288M}. There are also other multistage collapse scenarios \cite{2004AstL...30...14I,Bisnovatyi_Kogan_2017} that predict the presence of a neutrino pulse flux at longer times, $\sim 20\ s$.

In this regard, it is reasonable to study the algorithms for the joint registration of events in which bursts of neutrino pulses significantly correlate in time with the excitations of the gravitational detector. In fact, a similar algorithm was used in the processing of neutrino-gravity data during the SN 1987A flare \cite{Aglietta_1987,1991NCimC..14..171A,1991NCimB.106.1257A,ALEXEYEV1988209,2000JETP...91..845R}.

 The initial reference points will be considered the times $t_{k}\left( E\right)  $ of the appearance of "neutrino signals" of the BUST in the channel (measuring mode) "search for collapsars" at those observational monitoring intervals $\left( \Delta T=T_{2}-T_{1}\right)  \approx 20 \  s$, which fix an abnormally increased count rate " neutrino background” (see Section 4); the designation $t_{k}\left( E\right)  $ is adopted to emphasize also the possibility of estimating the energy release of the event $E$).
Further, at the same time intervals, the signals (emissions) of the OGRAN gravitational detector are studied.

Let us consider a separate (single) neutrino event (count) registered at the BUST at the time $t_{k}\left( E\right)  $ with energy release $E$. We assume that it can be accompanied by a delayed (or advanced) gravitational burst arising at the output of the gravitational detector optimal filter matched with the GW radiation model in the form of a train of duration $\tau_{s} $. 
Then the moment of appearance of the gravitational burst is estimated as
 $\tau_{k} =\left( t_{k}(E)+\tau_{s} +\delta \tau \right)  $, where $\delta \tau$ is some shift between neutrino and gravitational events, limited by $\left| \delta \tau \right|  \leq \Delta \tau $. The range of used shifts $\Delta \tau $ is set from physical considerations using astrophysical models of collapse \cite{1987ApJ...318..288M,Melson_2015,2004AstL...30...14I,Bisnovatyi_Kogan_2017}
  
Let's move on to the problem of detecting the correlation between the registered packets of neutrino - gravitational signals.
We use a model in which the gravitational detector output signal (after the optimal filter) is written as $z(t)=\vartheta S(t)+n(t)$, $S(t)=\sum_{k} s_{k}\left( t-\tau_{k} \right)  $ on the interval $T_{1}\leq t\leq T_{2}$;
the symbol $\vartheta =\left( 1,0\right)  $ is a formal parameter of the presence or absence of a signal.
This model assumes that the signal disturbance is given by the sum of individual non-overlapping pulses. An individual impulse in a burst $s_{k}(t)$ can be represented as (see Section 2)
$s_{k}(t)=a_{k}Re\left\{ exp\left( -j\varphi_{k} \right)  \widetilde{f}_{k} exp\left( -jw_{0}t\right)  \right\}  \ast g(t)$, for $0\leq t\leq \tau_{s} $.

The simplified version of the model also contains the assumption that the durations of individual pulses are equal, i.e. $\tau_{sk} =\tau_{s} $, and the universality of delay shifts between gravitational and neutrino signals. In addition, the considered neutrino-gravitational events are considered independent. Then the logarithms of the likelihood ratios of the individual events are added. It is natural to apply the maximum likelihood ratio criterion to our problem since its mathematical apparatus is well developed for detection problems against the background of Gaussian noise.

In particular, it is known that for narrowband normal noise (the OGRAN detector), the logarithm of the likelihood ratio $\Lambda (z)$ is proportional to the square of its output signal envelope $R(t)$ \cite{1975MIzSR....T....L}, 
$ln\Lambda_{k} (z) \propto R^{2}\left[ t_{k}\left( E\right)  +\tau_{s} \right]  $ - for one impulse; $ln\Lambda (z)=\sum_{k} ln\Lambda_{k} (z)$ for a pack.

The procedure for obtaining a solution (registration of the presence of a correlation) corresponds to finding the maximum of "sufficient statistics" $D $ - the integral likelihood ratio with a variation in the time shift between neutrino - gravitational events $\delta \tau$, i.e.
\begin{align}\label{eq17}
D=max\arrowvert_{\left| \delta \tau \right|  \leq \Delta \tau }\sum_{k=1}^{Q} \left\{ R^{2}\left[ t_{k}\left( E\right)  +\tau_{s} +\delta \tau \right]  /\sigma^{2} \right\}  \geq C_{\alpha }
\end{align}
here $Q$ is the number of pulses in the packet, $\sigma^{2}$ is noise dispersion at the output of the optimal filter, $C_{\alpha }$ is decision threshold $\vartheta =1$ with significance level $\alpha $, the calculation of which requires knowledge of the distribution function of sufficient statistics $D$.

Taking into account that with a Gaussian noise background the statistics of the variable $\sum_{k} R^{2}\left( \tau_{k} \right)  /\sigma^{2} $ at $\vartheta =0$ obeys the $\chi^{2} $ distribution, for which the integral probability density has the form $F_{1}\left( x\right)  =\Gamma \left( Q,x/2\right)  \Gamma \left( Q\right)  $, where $\Gamma \left(m, x\right)  $ is an incomplete gamma function;
using its explicit expressions, we find the following estimates
$F_{1}\left( C_{\alpha }\right)  \approx \left( 1-\alpha \right)^{\nu }  ,\nu \approx \Delta \tau \left( -{\rho }^{\prime \prime }_{0}  \right)^{1/2}  $, which make it possible to calculate the threshold $C_{\alpha }$ to confirm the neutrino-gravitational correlation according to the OGRAN and BUST data. 

\section{Conclusions }

A priori, the development of an algorithm for estimating the neutrino-gravitational correlation seems to be an incorrect task due to completely different principles (mechanisms of action) of the corresponding detectors. In this work, this can be done by reducing the output signals of the detectors to one type of random process - the Poisson pulse stream. The question of the presence of a statistical connection between two stochastic impulse flows is solved based on the "correlation criterion". The combined observable variable ("optimal" or "sufficient statistics") is chosen as the sum of gravitational signals arising in a close time neighborhood of neutrino events. In fact, such a variable is proportional to the cross-correlation function of the two considered Poisson processes. The correlation criterion corresponds to the maximum value of the observed variable in the space of mutual temporal shift of impulse processes of different physical nature. In this approach, the solution of the problem to the maximum, in addition to calculating the probability of the presence of a correlation, also gives an estimate of the delay of the neutrino signal relative to the gravitational one (flux shift corresponding to the maximum) and, thereby, an estimate of the rest mass of the neutrino (at the GW light speed).
Of course, the algorithm presented in this paper is a simplified scheme of the first approximation. It is limited by the assumptions that the considered Poisson fluxes are homogeneous (with a constant average pulse repetition rate) as well as the delay between neutrino and gravitational signals. The model in which the flow of neutrino signals has a time-varying average pulse repetition rate will apparently be more adequate to the physics of the BUST operation in the “search for collapsars” mode. However, this significantly complicates the calculation of the statistical characteristics of detection in the calculations of such a scheme. The authors are planning such a study in the next work.

The last remark is related to the degree of generality of the proposed search algorithm for neutrino gravitational correlation. In other words, can it be used for other more sensitive gravitational detectors such as LIGO/Virgo and neutrino telescopes such as Super-Kamiokande, DUNE, Hyper-Kamiokande, and JUNO. The answer, by and large, is no. Our development is directly related to a specific pair of OGRAN - BUST, since the optimal filtering of the output signals of the detectors depends on their principle of operation and design. But most importantly, the experiment under discussion - the search for collapses in the Galaxy - belongs to the category of "registration of rare events" - at best, one event in $30$ years. Such observations are usually not included in the priority programs of the aforementioned detectors.

Preference is given to significantly more probable events with a higher probability of occurrence. This is achieved by taking into account a large number of distant galaxies while increasing the radius of the observation sphere and, consequently, the distances to possible sources of gravitational and neutrino signals. In this case, registration of neutrino bursts with energies of the order of $10 \ MeV$ is impossible due to the insufficient sensitivity of scintillation detectors to cover very distant objects. They are being replaced by Cherenkov-type telescopes which include the installations mentioned above. The energies of recorded bursts from cosmic catastrophes (including special types of collapses) are in the range from $GeV$ to $TeV$ and higher. Examples of such registrations can be found in \cite{source_33} for the Ice Cube facility in Antarctica and \cite{source_34} for the Antares facility in the Mediterranean Sea. Registration of low-energy events with energies on the order of tens and hundreds of $MeV$ by the Ice Cube facility was discussed in \cite{source_35}.

Attempts to compare these events with data from LIGO gravitational detectors are contained in \cite{ANTARES}. Algorithms for searching for such correlations will apparently be refined. However, a common feature of the refined versions will be the optimization of the time setting of the shift interval between the gravitational and neutrino signals.

\section{Acknowledgments}
The authors are grateful to the directors of the SAI MSU Postnov K.A. and Cherepashchuk A.M. for interest in the problem and useful discussions, as well as to colleagues from the BNO INR RAS Petkov V.B., Novoseltseva R.V. for explaining the details of the physics of the BUST.

This research was funded by RFBR grant number 19-29-110-10.

\bibliographystyle{ieeetr}
\bibliography{Bib2.bib}

\end{document}